**Flexural wave illusion on a curved plate**

*Pengfei Zhao, Liyou Luo, Yongquan Liu, Jensen Li[*]*

P. Zhao, L. Luo, J. Li

Department of Physics, The Hong Kong University of Science and Technology,

Clear Water Bay, Kowloon, Hong Kong, 999077, P. R. China,

Y. Liu

State Key Laboratory for Strength and Vibration of Mechanical Structure,

School of Aerospace Engineering, Xi'an Jiaotong University, Xi'an, 710049, China

* E-mail: jensenli@ust.hk

Abstract

Manipulating elastic waves using a transformation approach is challenging due to the complex constitutive relationship. However, for flexural waves, approximated as scalar waves, two straightforward approaches emerge based on geometric curvature and plate thickness. Here, we develop transformation theory to establish equivalence between curved plates of different shapes and thickness profiles. By introducing tailor-made thickness profiles on a given curved shape enables illusion effects, where flexural waves propagate as if on a flat plate or on another curved plate with totally different configuration. Numerical simulations and experimental field mapping confirm the effectiveness of these illusions. Our approach on flexural wave illusion finds applications in structural designs with material and shape constraints, and holds potential for vibration control, wavefront shaping, chaotic dynamics and topology control.

Keywords: illusion effect, curved plate, inhomogeneous elastic material, transformation elastodynamics

# 1. Introduction

Over the past two decades, transformation optics (TO) has emerged as a valuable tool for designing various optical devices. It has been successfully applied to create invisibility cloaks in electromagnetic waves[1-9] and later extended to other classical waves such as acoustic waves[10-14], water waves[15-17], and elastic waves in solids[18-20]. Additionally, transformation or related approaches have been utilized for illusion optics in drawing equivalence for optical appearance between two distinct configurations, e.g. by adding an additional permittivity profile surrounding, being external to, or on top of an object.[21-24] However, in the realm of elastic waves in solids, the transformation approach is more challenging due to the complex elastic constitutive relationship or the lack of form invariance upon a coordinate transformation.[25,26] Recent attention has focused on developing complex elastic materials, such as pentamode, micropolar and Willis metamaterials, to overcome these challenges and enable more accurate transformation-based designs.[27-33]

On the other hand, when considering a special type of elastic waves known as flexural waves, their complexity is reduced as they can be approximated as scalar waves consisting primarily of the out-of-plane component. This approximation suggests that local dispersion, represented by a refractive index ellipsoid, can be used to manipulate these waves. One approach is to utilize intrinsic geometric curvature, from a curved plate of constant thickness of a single material, to refract the waves.[34,35] Recently, such a pure curvature approach has been used to realize different gradient-index lenses with specific refractive index profiles, such as a Luneburg lens with cylindrical geometry.[36] Interestingly, without altering the material composition of the plate, the local change of refractive index can also be achieved by varying the thickness profile of the plate. This technique has been employed to achieve signal isolation and also different gradient-index lenses.[37-39] Considering that both intrinsic geometric curvature and plate thickness can modify the local dispersion relationship of flexural waves, their interplay and combination of them are expected to provide a versatile platform for manipulating flexural waves, particularly in the aforementioned applications enabled by the transformation approach.

In this work, we develop a transformation theory and demonstrate its applications in illusion elastodynamics. Firstly, for a curved plate (or shell) with a given shape, we solve Liouville's equation in the physical domain to achieve the "flattening" effect. By introducing a thickness profile, we compensate for the curvature, allowing a flexural wave to propagate on the curved plate as if it were flat. Secondly, we extend the transformation theory to create illusions by instructing a thickness profile on a given curved shape to mimic another curved plate with totally different configuration. The effectiveness of flattening and illusion effects is validated through ray tracing, full-wave simulations, and field mapping experiments. The ability to establish equivalence between different curved plates has significant implications in vibration control, aerospace, and mechanical engineering, particularly when we have constraints on material consumptions and shape considerations for designing various mechanical structures.[40,41]

## 2. Theory

To begin, let us consider a special case of an illusion where our objective is to transform a curved plate with a given shape function $z(x, y)$ into a flat plate with a uniform thickness $h_0$, as shown in **Figure 1**(a), which we refer to as "flattening." Normally, a flexural wave traveling on the curved plate would experience scattering due to the curvature.[34] As a preliminary step towards a comprehensive version of illusion, we introduce a tailor-made thickness profile, h(x, y), with the aim of eliminating this scattering effect. Consequently, a flexural wave traveling on the curved plate in physical space, defined by coordinates (x, y), behaves as if it were on a flat plate in the virtual space, defined by coordinates (u, v), in the language of transformation optics. With reference to the flat plate of the same material, the flexural wave travels with a refractive index profile, which is independent of the material parameters. It is denoted as $n(x, y) = \sqrt{h_0/h(x, y)}$ in Kirchhoff-Love plate theory[38], contributing to the overall metric as $n(x, y)^2(dx^2 + dy^2 + dz^2) = du^2 + dv^2$. The overall metric tensor is written as

$$g_{\text{all}}(x, y) = n(x, y)^2 g_s(x, y) = n(x, y)^2 \begin{pmatrix} 1 + (\partial_x z)^2 & (\partial_x z)(\partial_y z) \\ (\partial_x z)(\partial_y z) & 1 + (\partial_y z)^2 \end{pmatrix}. \tag{1}$$

Here, $g_s(x, y)$ refers to the metric tensor of the surface part without the index profile. In fact, the overall Gaussian curvature of the metric tensor gall, denoted as $K_{\text{all}}$, remains invariant upon coordinate transformation and must be zero due to the Euclidean nature of the virtual space. By applying the Brioschi formula to evaluate $K_{\text{all}}$ from the components of $g_{\text{all}}$, we obtain

$$n^2 K_{\text{all}} = K[z] - \Delta[g_s](\ln n) = 0, \tag{2}$$

where $n(x, y)$ can now be solved in the physical space by applying the Dirichlet boundary condition $\ln(n) = 0$ at the outer boundary of the device ($|x| = L$ or $|y| = L$) in this work. This boundary condition is made under the consideration that the curved plate is embedded within a background flat plate of uniform thickness $h_0$. It is worth noting that devices with shapes other than a square are also possible. Here, $K[z]$ represents the Gaussian curvature of the surface $z(x, y)$ without the index profile and is evaluated in the physical space by

$$K[z] = \frac{(\partial_x^2 z)(\partial_y^2 z) - (\partial_x \partial_y z)^2}{\left(1 + (\partial_x z)^2 + (\partial_y z)^2\right)^2},$$

and $\Delta[g]$ is the Laplacian-Beltrami operator with respect to a specific metric $g$. It can be evaluated in the physical space by

$$\Delta[g](\circ) = \frac{1}{\sqrt{\det g}} \begin{pmatrix} \partial_x & \partial_y \end{pmatrix} \left( \sqrt{\det g} \, g^{-1} \begin{pmatrix} \partial_x \\ \partial_y \end{pmatrix} \circ \right).$$

Additionally, we recognize that Equation (2) is known as Liouville's equation, which is typically expressed as a nonlinear partial differential equation in isothermal coordinates $(u, v)$. However, in our case, we transform and solve it in terms of $(x, y)$ coordinates as a linear partial differential equation of $\ln n$.

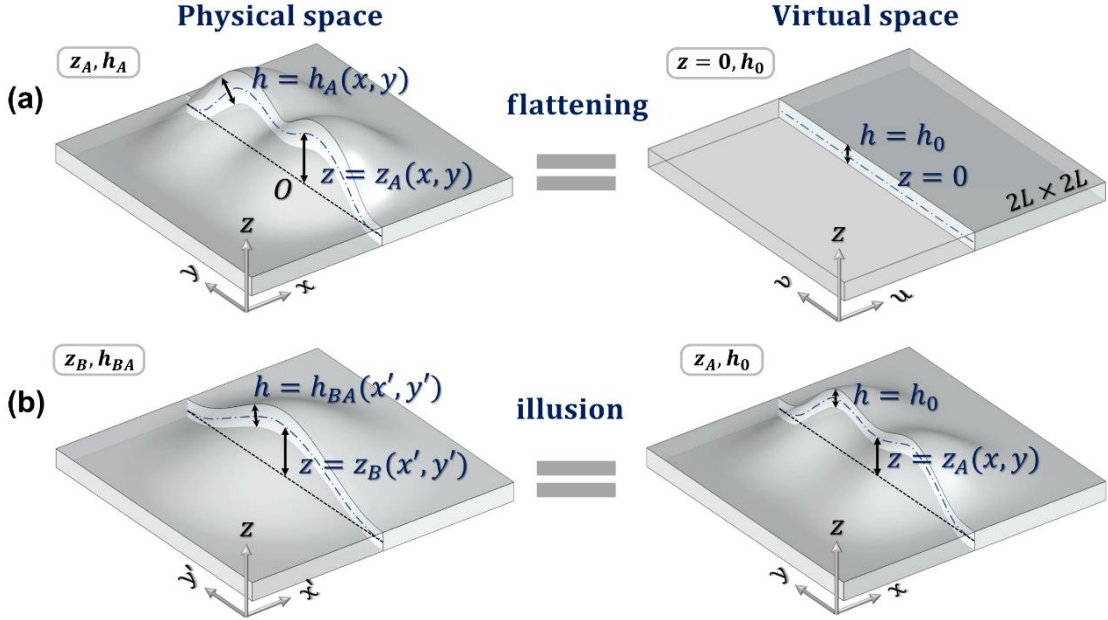

**Figure 1.** (a) A curved plate of given shape $z = z_A(x, y)$ is "flattened" by a tailor-made thickness profile $h = h_A(x, y)$ in the physical space to a flat plate ($z = 0$) of uniform thickness $h_0$ for flexural wave propagation in the virtual space $(u, v)$. (b) A curved plate of given shape $z = z_B(x', y')$, with a tailor-made thickness profile $h = h_{BA}(x', y')$ in the physical space, is equivalent to $z = z_A(x, y)$ with constant thickness $h_0$ in the virtual space as an illusion effect.

On the other hand, the Euclidean nature of the virtual space when interpreted in physical space also implies $\Delta[g_{\text{all}}](u(x, y)) = \Delta[g_{\text{all}}](v(x, y)) = 0$. As Equation (1) gives $\Delta[g_{\text{all}}](\circ) = n^{-2}\Delta[g_s](\circ)$, we can determine the coordinate transformation $u(x, y)$ and $v(x, y)$ by solving

$$\Delta[g_s](u(x, y)) = 0, \tag{3}$$

$$\Delta[g_s](v(x,y)) = 0,$$

in the physical space again with Dirichlet boundary conditions $u(x,y) = x$ and $v(x,y) = y$ at the outer boundary of the device. For convenience, we summarize the transformation between the two spaces when flattening a given surface $z = z_A(x,y)$ using the thickness profile $h = h_A(x,y)$ as follows:

$$\{z_A, h_A\} \leftrightarrow \{z = 0, h_0\} \text{ with } (x,y) \leftrightarrow (u,v), \quad (4)$$

while $h_A(x,y)$, $u(x,y)$ and $v(x,y)$ are solved independently from Equation (2) (expressing $n$ in terms of $h$) and Equation (3) in our approach. While our primary objective is to solve for $h_A(x,y)$ without the need to explicitly solve for $u(x,y)$ and $v(x,y)$ for implementation of the flattening, obtaining the coordinate transformation facilitates the subsequent step towards achieving a more general illusion.

Suppose we have a given shape $z_B(x',y')$ and we would like to introduce a tailor-made thickness profile $h_{BA}(x',y')$ so that it is equivalent to the previous curved plate before flattening, i.e. $z_A(x,y)$ with a constant thickness $h_0$, as an illusion effect shown in Figure 1(b). We define the associated problems in flattening $z_A$, listed in Equation (4), and in flattening $z_B(x',y')$ with $h_B(x',y')$, listed as

$$\{z_B, h_B\} \leftrightarrow \{z = 0, h_0\} \text{ with } (x',y') \leftrightarrow (u,v). \quad (5)$$

After obtaining $h_A$, $h_B$ and the two coordinate transformations, we then cascade the two transformations to arrive

$$\{z_B, h_{BA}\} \leftrightarrow \{z_A, h_0\} \text{ with } (x',y') \leftrightarrow (x,y), \quad (6)$$

where the thickness profile $h_{BA}$ should now be implemented according to

$$h_{BA}(x',y') = h_0 h_B(x',y')/h_A(x,y). \quad (7)$$

In this case, $(x',y')$ is the physical space for realization of the illusion effect while $(x,y)$ is the virtual space in drawing equivalence. In deriving Equation (7), we have multiplied a common profile $h_0/h_A$ to the thickness profiles at the corresponding coordinates for configurations A and B after combining Equation (4) and (5). In other words, when we multiply $\sqrt{h_A/h_0}$ to the index profile in the $(x,y)$-space, the same $\sqrt{h_A/h_0}$ profile has to be multiplied to the index profile in the $(x',y')$-space according to transformation optics theory.[1,2]

## 3. Experimental demonstration of illusion effect for flexural waves

As a specific type of illusion effect, we consider flattening a curved plate with a shape function defined by

$$z_A(x,y) = z_0(1-\tilde{x}^2)^4(1-\tilde{y}^2)^4(1+\alpha\tilde{x}^2)(1+\beta\tilde{y}^2), \quad (8)$$

where $\tilde{x} = x/L$, $\tilde{y} = y/L$ range from $-1$ to $1$. For $L = 150$ mm, $z_0 = 40$ mm, $\alpha = 1/2$, and $\beta = 10$, the surface has a double hump shape, as shown as the inset of **Figure 2**(a). The thickness profile within the curved region, shown in the inset, ranges from 2.9 mm (at four locations around the two humps) to 4.9 mm (near the peaks of the two humps). First, we performed ray tracing on the curved plate for the configuration with a tailor-made thickness profile $h_A(x, y)$, by solving Equation (2), in the "geometrical optics" limit. A point source was placed at $(x, y) = (0, 250$ mm$)$ within a background flat plate of the same thickness outside the curved region. The results are shown in Figure 2(a) in red for the rays emitted from the point source with exiting angles in steps of $10°$. In the current case, labelled as $\{z_A, h_A\}$, the rays bend around the humps and exit in their original direction as if they travelled from the original point source in straight lines, effectively flattening the surface. Within the curved region, a regular grid of virtual coordinates $(u, v)$, in steps of 15 mm, is drawn as yellow lines. Although the grid drawn in $(x, y)$ coordinates is distorted near the two humps, the rays appear to travel in "straight lines" in terms of the virtual coordinates. In contrast, for the configuration $\{z_A, h_0\}$ with the same shape but with constant thickness $h_0 = 3$ mm. The results are shown in Figure 2(b). As the rays enter the curved region (dashed square of size $2L \times 2L$), the curvature deflects them, causing some to cross each other and focus in the region at the bottom side of the curved region.

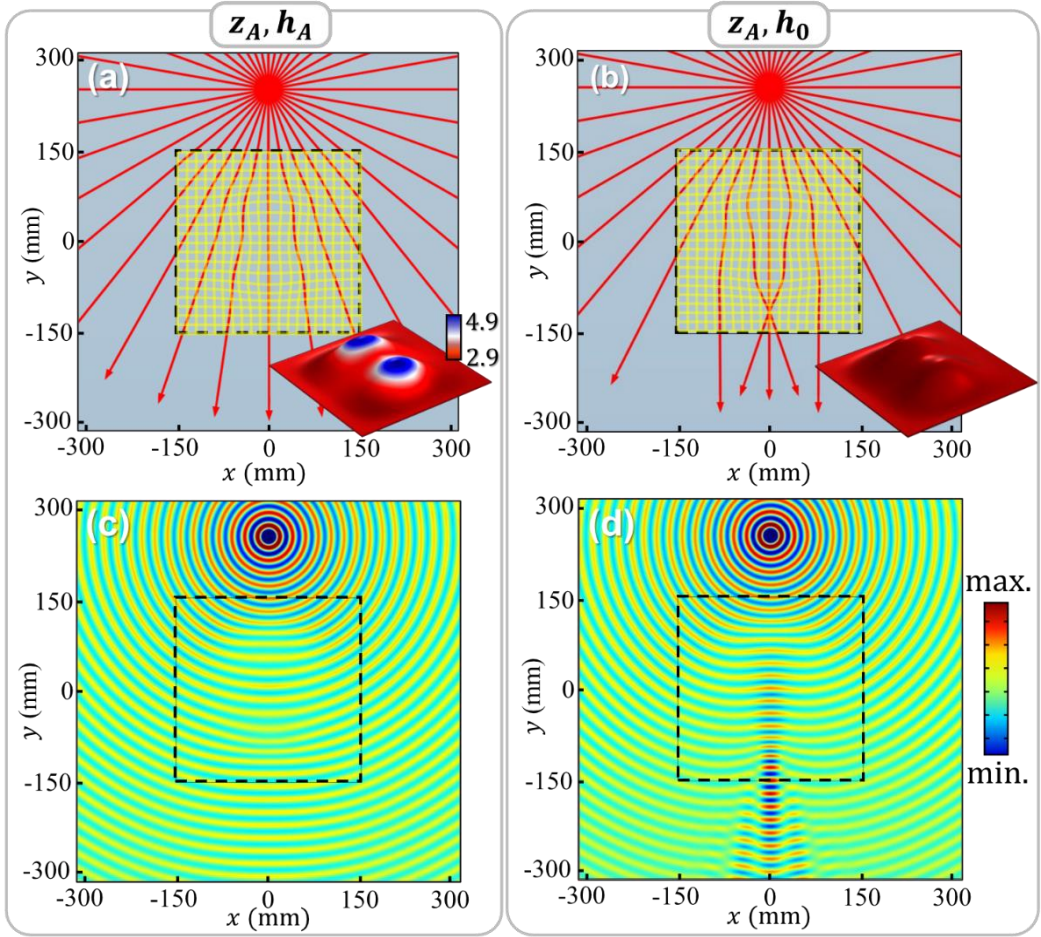

**Figure 2.** Ray tracing for a point source located 100 mm away from the top side of the curved region for the configurations (a) $\{z_A, h_A\}$ for inhomogeneous thickness profile for flattening and (b) $\{z_A, h_0\}$ with a constant thickness $h_0 = 3$ mm for comparison. The corresponding out-of-plane velocity profile (real part) of full-wave simulation for a flexural wave point source at the same location with a frequency of 20 kHz are shown in (c) and (d) respectively. The insets illustrate the curved surfaces with the thickness profiles (in mm) being investigated.

As illustration, we selected a frequency of 20 kHz and performed the corresponding full-wave simulations (using COMSOL Multiphysics) for both the $\{z_A, h_A\}$ and $\{z_A, h_0\}$ configurations, as shown using the real part of the out-of-plane velocity projected on the x-y plane in Figure 2(c) and (d) respectively. The results confirm the previous ray-tracing results with additional insights. For the $\{z_A, h_A\}$ configuration, where the double hump is flattened by introducing $h_A$, the wavefront smoothly passes through the curved region (with only geometric distortion according to the grid of virtual coordinates) and exits the curved region with an almost circular wavefront and amplitude that is not sensitive to the direction from the point source. On the other hand, for the $\{z_A, h_0\}$ configuration, the focusing effect is evident as a beam of higher intensity at the exit of curved region. As the field further propagates, dislocations in phase become apparent when the field develops two side lobes.

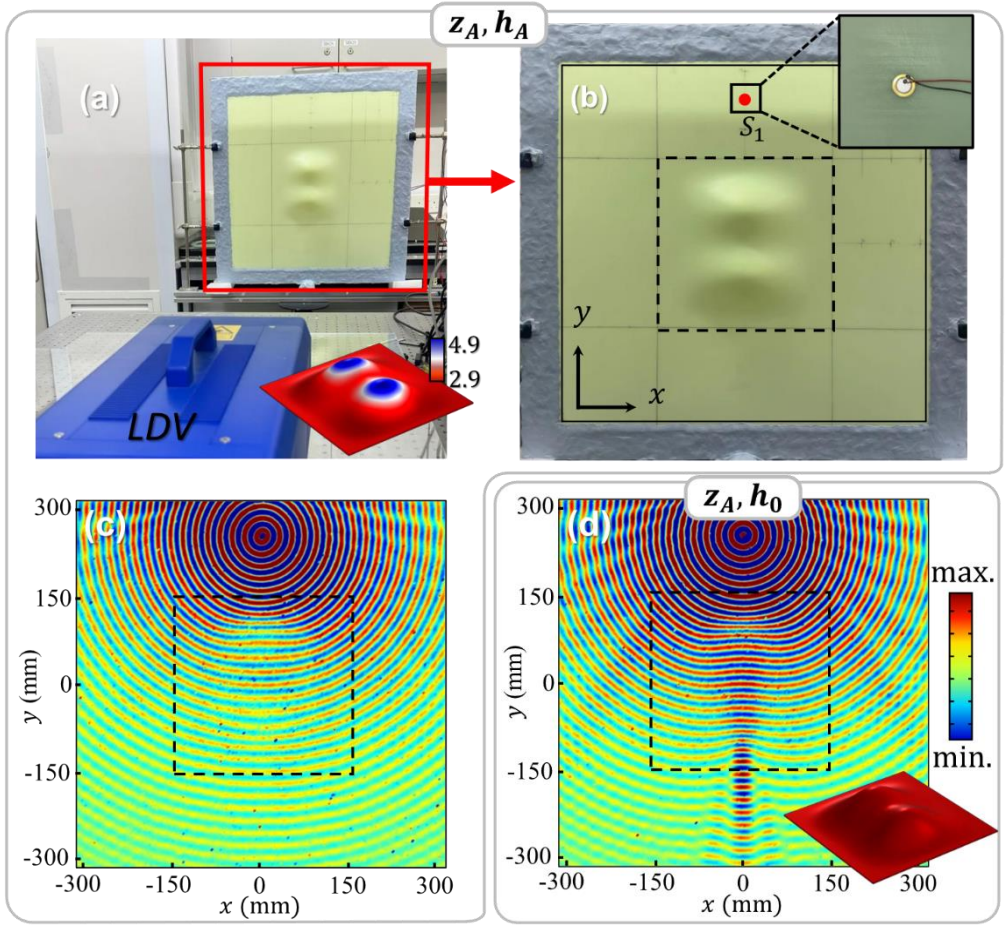

**Figure 3.** (a) Experimental setup including the sample $\{z_A, h_A\}$ and a scanning LDV. (b) A photograph of the sample with the (firing) transducer at $S_1$, 100 mm away from the top edge of the curved region, at the back side shown in its inset. (c) Experimental results (out-of-plane velocity field) for configuration $\{z_A, h_A\}$ with tailor-made thickness profile. (d) Experimental results for the configuration $\{z_A, h_0\}$ with constant thickness $h_0 = 3$ mm for comparison.

**Figure 3**(a) shows the experimental setup. A scanning laser Doppler vibrometer (OptoMET, marked as LDV) was used with the laser directed perpendicular to the surface. In the current work, samples are fabricated by 3D printing using a cured photosensitive resin, with density $\rho_0 = 1190 \text{ kg} \cdot \text{m}^{-3}$, Young's modulus $E_0 = 3.4$ GPa, and Poisson ratio $\nu = 0.35$. Figure 3(b) shows the sample fabricated for the configuration $\{z_A, h_A\}$, with the dashed square (300 × 300 mm²) enclosing the curved region $z_A$, which is further embedded in a background flat plate of constant thickness $h_0 = 3$ mm. A piezoelectric transducer with a diameter of 12 mm was positioned 100 mm away from the top edge of the curved region at the backside, referred to as $S_1$. Stripes of Blu Tack were adhered to the outer boundary of the entire sample (760 × 760 mm$^2$) to minimize reflection from the boundary, leaving out a measurement domain outlined by a solid square of size 620 × 620 mm$^2$. The out-of-plane velocity field is then mapped point-by-point by the LDV at a frequency 20 kHz, as shown in Figure 3(c), for the configuration $\{z_A, h_A\}$ with the tailor-made thickness profile $h_A(x_1, y_1)$ applied to the surface $z_A$ for flattening. As expected, a circular wavefront emerges from the exit side of the curved region, indicating good agreement with the simulation result presented in Figure 2(c). We note that some additional angular fringes are visible in the amplitude, particularly at large angles away from the curved surface. These fringes are attributed to imperfect absorbing boundaries of the sample, leading to some residue reflections from the boundary. Figure 3(d) shows the experimental results when the curved surface has only curved shape $z_A$ and a constant thickness, representing the $\{z_A, h_0\}$. The field becomes focused at the bottom side of the measurement domain. The wavefront experiences significant distortion, resulting in amplitude fluctuations at the exit side of the bottom. The result serves as a control experiment.

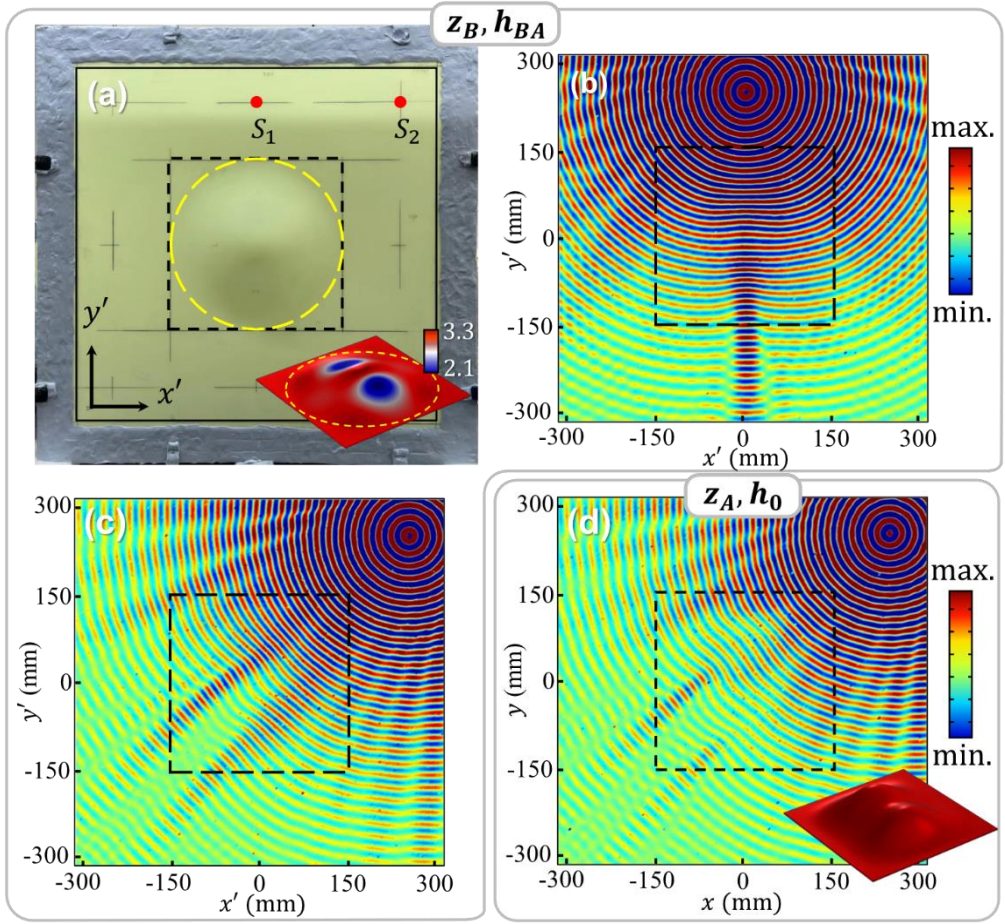

**Figure 4.** (a) A photograph of the sample $\{z_B, h_{BA}\}$ featuring a cosine shape to mimic another configuration $\{z_A, h_0\}$ (the sample in Figure 3(d)), with the (firing) transducer put at either $S_1$ or $S_2$ (250 mm to the right of the $S_1$). (b) Experimental results (out-of-plane velocity field) for the point source at $S_1$ for configuration $\{z_B, h_{BA}\}$. (c) Experimental results obtained for the point source at $S_2$ for configuration $\{z_B, h_{BA}\}$. (d) Experimental results for the point source at $S_2$ for configuration $\{z_A, h_0\}$ for comparison. The insets illustrate the thickness profiles for each sample.

Now, we consider the illusion between two different curved plates. As an example, we choose a cylindrical symmetric shape by $z_B(x', y') = z_0 \cos^2(\pi r'/(2L))$ for radius $r' = \sqrt{x'^2 + y'^2}$ ranging from 0 to $L$, and $z_B$ returns to $z = 0$ for $r'$ larger than $L$. The values of $L$ and $z_0$ remain the same as before. Our goal is to mimic the configuration $\{z_A, h_0\}$ (Equation (8)) by introducing a tailor-made thickness profile $h_{BA} = h_0 h_B/h_A$. This thickness profile is constructed from $h_A$ and $h_B$, which flatten $z_A$ and $z_B$ from Equation (4) and (5), respectively. **Figure 4**(a) shows a photograph for the sample configuration $\{z_B, h_{BA}\}$. In addition to the point source at the same location $S_1$ as before, another possible location $S_2$, positioned 250 mm to the right of $S_1$, is also indicated. The shape of $z_B$ and the corresponding thickness profile $h_{BA}$, ranging from 2.1 to 3.3 mm, are shown in the inset of Figure 4(a). Essentially, the lack of cylindrical symmetry for the shape $z_A$ can now be equivalently implemented by a cylindrically symmetric $z_B$ but with a thickness profile that lacks such symmetry. Figure 4(b) shows the experimentally measured out-of-plane velocity field for the configuration $\{z_B, h_{BA}\}$ when a point source is activated at $S_1$ at a frequency of 20kHz. In the region outside the curved region (dashed square), the field exhibits a focusing effect similar to the results for the configuration $\{z_A, h_0\}$, previously shown in Figure 3(d). It is important to note that the illusion effect does not depend on the location of the point source. Figure 4(c) shows the experimental results when the point source is activated at $S_2$ at the same frequency for the current configuration $\{z_B, h_{BA}\}$. The wavefront leaves the curved surface with two clearly visible focusing beams around the bottom left corner. The field pattern outside the dashed square should resemble the results for the equivalent configuration $\{z_A, h_0\}$. In fact, we have conducted an experiment for a point source at $S_2$ on the configuration $\{z_A, h_0\}$, with measured velocity field profile confirming the resemblance, as shown in Figure 4(d). It is worth noting that the configuration $\{z_B, h_{BA}\}$ actually uses less materials than the configuration $\{z_A, h_0\}$ and has a confined shape of cylindrical symmetry. These aspects can be beneficial for optimizing structural designs with constraints in material consumptions and shape considerations [40].

## 4. Conclusion

In conclusion, we have formulated and experimentally demonstrated a transformation theory for flexural wave illusion using curved plates with inhomogeneous thickness profiles. The ability to establish equivalence between curved plates of different shapes and thickness profiles allows us to have flexible control over the design process. This enables us to prioritize either the shapes (curvature) or the materials (thickness profile) to achieve the desired propagation phenomena (prescribed scattering behavior or the flattening of a curved plate into a flat plate). The developed transformation approach will be useful for vibration control, structural design optimization, and can be extended to other types of waves such as electromagnetic waves. The curved space approach, enabled by the current transformation ability, can also be used to control chaotic dynamics and topology.[42,43]

## 5. Experimental Section

*Experimental Setup*: A 3D-printer (UnionTech, Lite800) had been used to fabricate the samples of Figure 3(b) and Figure 4(a). The material of the samples using a cured photosensitive resin has the elastic parameters: mass density $\rho_0 = 1190 \text{ kg} \cdot \text{m}^{-3}$, Young's modulus $E_0 = 3.4$ GPa, and Poisson ratio $\nu = 0.35$. Piezoelectric diaphragms (with diameter of 12 mm) located on the $S_1$ and $S_2$ were used to excite the harmonic flexural wave at a frequency 20 kHz on the plate. The boundaries of the entire samples ($760 \times 760$ mm$^2$) were covered with stripes of Blu-Tack to reduce the reflection, leaving out a measurement domain of $620 \times 620$ mm$^2$. A scanning laser Doppler vibrometer (OptoMET, marked as LDV) was used with the laser directed perpendicular to the surface for measuring the out-of-plane velocity field distribution.

*Numerical Simulations*: All 3D full-wave simulations were conducted with the finite element software COMSOL Multiphysics. Perfectly-matched layers were set on the outer boundaries of all samples in simulation. The largest mesh-element was set to be smaller than half of the smallest thickness profile of the plate.

**Acknowledgment**

P. Z. and L. L. contributed equally to this work. J.L. acknowledges support from Research Grants Council (RGC) of Hong Kong through projects no. 16307522 and AoE/P-502/20. Y.L. acknowledges support from the National Natural Science Foundation of China (Grant No. 12172271).

**Conflict of Interest**

The authors declare no conflict of interest.